# 20 YEARS OF SPELEOTHEM PALEOLUMINESCENCE RECORDS OF ENVIRONMENTAL CHANGES: AN OVERVIEW

Yavor Y. Shopov

**Abstract**
This paper discusses advance of the research on Speleothem Paleoluminescence Records of Environmental Changes after it have been first introduced by the author 20 years ago. It is demonstrated that most of the progress in this field was made in result of the operation of the International Program "Luminescence of Cave Minerals" of the commission on Physical Chemistry and Hydrogeology of Karst of UIS of UNESCO.
Potential, resolution and limitations of high resolution luminescence speleothem proxy records of Paleotemperature, Solar Insolation, Solar Luminosity, Glaciations, Sea Level advances, Past Precipitation, Plants Populations, Paleosoils, Past Karst Denudation, Chemical Pollution, Geomagnetic field and Cosmic Rays Flux variations, Cosmogenic Isotopes production and Supernova Eruptions in the Past, Advances of Hydrothermal Waters, and Tectonic Uplift are discussed.
It is demonstrated that speleothems allow extremely high resolution (higher than in any other paleoclimatic terrestrial archives) and long duration of records. Some speleothems can be used as natural climatic stations for obtaining of quantitative proxy records of Quaternary climates with annual resolution.

## Introduction

Luminescence is the most sensitive to depositional conditions property of cave minerals (Tarashtan, 1978). Therefore it can be used for reconstruction of these conditions. So in 1988 commission of Physical Chemistry and Hydrogeology of UIS of UNESCO decided to start an international programme on study of "Luminescence of Cave Minerals". Most of the further progress in the field was made in result of the operation of this programme. Infact it developed a new fielf of science called "paleoluminescence" like paleomagnetism.
Many speleothems exhibit luminescence when exposed to ultraviolet (UV) light sources or other high-energy beams. In dependence of the excitation source there are specific kinds of luminescence: "photoluminescence" (excited by UV and other light sources), "X-ray luminescence" (by x-rays), "Cathodoluminescence" (by electron beam), "Thermoluminescence" (by heat), "Candoluminescence" (by flames) and "triboluminescence" (by crushing). Different types of excitation may excite different luminescent centers- electron defects of the crystal lattice; admixture ions substituting ions in the crystal lattice or incorporated in cavities of that lattice; inclusions of other minerals; or fluid

## Table 1 - Origin of luminescence of Cave Minerals

| Luminescence activator | Excitation source | Color of luminescence | Afterglow | Origin | Reference |
|---|---|---|---|---|---|
| **Calcite** | | | | | |
| 24 types of luminescence (see Shopov, 2004, same volume) | | | | | |
| **Aragonite** | | | | | |
| 11 types of luminescence (see Shopov, 2004, same volume) | | | | | |
| **Vaterite** | | | | | |
| 36. Organics? | $N_2$-Laser | blue-green | long | infiltration | Shopov (1989b) |
| 37. Organics? | $N_2$-Laser | green | long | infiltration | Shopov (p.c.) |
| **Huntite** | | | | | |
| 38. Organics? | $N_2$-Laser | blue | long | infiltration | Shopov (1988) |
| 39. Organics? | $N_2$-Laser | yellow-green | long | infiltration | Shopov (1988) |
| **Hydromagnesite** | | | | | |
| 40. Organics? | $N_2$-Laser | green | long | infiltration | Shopov (1989b) |
| 41. Organics? | $N_2$-Laser | yellow-green | long | infiltration | Shopov (1989b), |
| **Dolomite** | | | | | |
| 42. Organics | $N_2$-Laser | yellow-green | long | infiltration | Shopov (1989b) |
| **Gypsum** | | | | | |
| 43. Organics | $N_2$-Laser | yellow-green | long | infiltration | Shopov (1989b) |
| 44. Organics | Ar-Laser | yellow | long | infiltration | Shopov (1989b) |
| 45. ? | Hg (LWUV) | deep-yellow | long | infiltration | White, Brennan (1989) |
| 46. ? | Hg (LWUV) | orange | | infiltration | White, Brennan (1989) |
| 47. $Mn^{2+}$ | Ar –Laser | red | 0.1 s | hydrothermal | Shopov & al. (1989) |
| 48. $Fe^{3+}$ | fl. Ar -Laser | dark red+ IR | | hydrothermal | Shopov & al. (1989) |
| **Halite** | | | | | |
| 49. Organics | fl. Ar -Laser | blue | long | infiltration | Shopov (p.c.) |
| 50. Organics | fl. Ar -Laser | blue-green | long | infiltration | Shopov (p.c.) |
| **Darapscite** | | | | | |
| 51. Organics | fl. Ar -Laser | blue-green | long | infiltration | Shopov (p.c.) |
| **Purpurite** | | | | | |
| 52. ? | Ar -Laser | Green-yellow | | guano on lava | Shopov (1989b) |
| **$CaCO_3$-II** | | | | | |
| 53. $Pb^{2+}$ | Xe-lamp (SWUV) | UV | | infiltration | Shopov (1989b) |
| 54. Organics | Xe (SWUV) | violet | long | low-temp.h-t | Shopov (1989b) |
| 55. $Mn^{2+}$ | Ar -L, He-Ne-L | dark red | 0.1 s | hydrothermal | Shopov & al. (1988b) |
| 56. $Fe^{3+}$ | Xe-lamp (SUV) | dark red-IR | | hydrothermal | Shopov & al. (1988b) |
| **Opale** | | | | | |
| 57. Organics | Ar-Laser | yellow-green | 0.1s | ? | Shopov & Slacik (p.c.) |
| 58. Organics | Ar-Laser | yellow | 0.1s | ? | Shopov & Slacik (p.c.) |
| **Quartz** | | | | | |
| 59. $AlO_4^{-}$ | Xe-lamp (LWUV) | blue | | | Shopov (1989) |
| 60. $O^*_{Fe}$ | Xe-lamp (LWUV) | yellow | | | Shopov & al.(1989) |
| 61. $Fe^{3+}$ | Xe-lamp (SWUV) | dark red | | hydrothermal | Shopov & al.(1989) |
| **Hydrozincite** | | | | | |
| 62. Organics | $N_2$-Laser | yellow-green | | ore-weathering | Shopov (1989b) |
| 63. Organics | $N_2$-Laser | yellow | | ore-weathering | Shopov (1989b) |

continued

**Comments to table 1:**
*36- 41 Organic origin of luminescence of these minerals is under question, because molecules of absorbed water may cause similar luminescence (Tarashtan, 1978) in aggregates like studied samples (moonmilk).

Table 1 (follows) - Origin of luminescence of Cave Minerals
Luminescence of Cave minerals at other excitations

| Luminescence activator | Excitation source | Color of luminescence | Afterglow | Origin | Reference |
|---|---|---|---|---|---|
| **Cathodoluminescence** | | | | | |
| **Calcite** | | | | | |
| 64. $Mn^{2+}$ | electron beam | pink | | ? | White (1974) |
| 65. $CO_3^{3-}$ | electron beam | blue | | infiltration | Shopov (p.c.) |
| **Halite** | | | | | |
| 66. $Cl^-$ | defect, el. beam | violet | | infiltration | Shopov (p.c.) |
| **Darapskite** | | | | | |
| 67. ? | electron beam | violet | | infiltration | Shopov (p.c.) |
| **Candoluminescence** | | | | | |
| **Gypsum** | | | | | |
| 68. $Mn^{2+}$ | hydrogen flame | green | | | Hess et al (1971) White & Brennan (1989) |
| 69. ? | hydrogen flame | lemon yellow | | infiltration | White & Brennan (1989) |
| **Thermoluminescence** | | | | | |
| **Calcite** | | | | | |
| 70. ? | UV+ warming | glow at 105 K | | infiltration | White & Brennan (1989) |
| 71. ? | Xe+ heating | glow at 350-500 K | | ? | Dublyansky (in press) |
| **X-ray luminescence** | | | | | |
| **Calcite** | | | | | |
| 72. $CO_3^{3-}$ | X-rays | blue | | infiltration | Shopov & Spasov (1983) |

inclusions, molecules, ions or radicals adsorbed inside of the lattice (Shopov, 2004). Some or all of them may exist in a single speleothem Shopov (1997).

If the emission proceeds only during the excitation than it is called "fluorescence", if it proceeds after the termination of the excitation than it is called "phosphorescence". Some luminescent centers produce only fluorescence, but other both fluorescence and phosphorescence. The type of luminescent centers determines the colour of luminescence. Colour may vary with changes of the excitation sources.

Brightness of luminescence is function of the concentration of luminescence centers. It is almost linearly proportional to concentration of luminescent centers.

**Origin of luminescence of Cave Minerals**

Most known luminescent centers in calcite are inorganic ions of Mn, Tb, Er, Dy, U, Eu, Sm and Ce (Tarashtan, 1978, Shopov, 1986, Shopov et al., 1988). Luminescence of minerals formed at normal cave temperatures (below 40° C) is due mainly to molecular ions and adsorbed organic molecules. Luminescence of uranil- ion is also very common in such speleothems. Usually several centers activate luminescence of the sample and the measured spectrum is a sum of the spectra of two or more of them. Before using a speleothem for any paleoenvironmental work it is necessary to determine that all luminescence of the sample is due to organics (Shopov, 2004).

**Measurements and Photography of Luminescence**

The most common method of excitation is irradiation by UV light sources producing photoluminescence and when luminescence is usually spoken about it is with this kind of excitation in mind. Phosphorescence of speleothems in caves can be seen by irradiating of speleothems with a photographic flash with closed eyes, with following rapid opening of the eyes after flashing. Such "Visual Luminescent Analysis" (VLA) has been widely used in caves. For this purpose (Tarcus, 1981) used also other simple devices such as portable UV lamps with short wave UV (SWUV) and long wave UV (LWUV). Slacik (1977) used a simple apparatus, which registered total emitted light by a galvanometer with a photocell for the quantitative evaluation of the luminescence intensity. But data obtained by the VLA method are subjective and the determination of luminescence activators is not possible. In fact attempts to determine activators of the luminescence with VLA and chemical analysis lead to incorrect results.

Investigations of the spectra of luminescence reveal new possibilities for luminescent research in mineralogy: - for example determination of luminescent centers, the character of isomorphic substitution, structural characteristics of the admixtures and defect centers, and typomorphic peculiarities of minerals (Tarashtan, 1978).

Luminescence spectra of cave minerals have been measured by means of exciting them with nitrogen Lasers (Ugumori, Ikeya, 1980; Shopov, Spasov, 1983; Shopov, 1988, 1989a,b), Xe- or Hg-lamp (Shopov et al., 1988; White, Brennan, 1989), Argon Lasers (Shopov, 1989b, White, Brennan, 1989) or by He- Ne Lasers. A disadvantage of this method is that it is destructive and gives total spectra of luminescence of the entire sample and is inapplicable for research of fine mixed aggregates such as moonmilk.

The conventional method for photography of fluorescence (PF) is not adaptable for cave photography, because it needs long exposition times (30-60 min) and a permanent electric source. This method always distorts the colour of luminescence, because it is impossible to choose a pair of filters, which can absorb whole emission of the UV-lamp, without absorbing a part of the luminescence of the sample. If an UV- absorption filter is not used to absorb lamp emission, UV photos will be obtained instead of photos of fluorescence.

The simplest method for luminescent research (table 2) is Impulse Photography of Phosphorescence (IPP) (Shopov and Grinberg, 1985b, Shopov, 1989a, 1991). Equipment used for this consists of a photo camera with a shutter delayer, which opens the shutter, several milliseconds after flash emission ends. It uses ordinary photoflash to excite speleothem luminescence. Adding of an impulse UV-source (flash with an UV-passing filter) (IPFP, table 2) can give both photos of fluorescence and phosphorescence together or separately. Photo slides obtained by this method can be measured by Colour Slide Spectrophotomerty (CSS), (table 2) to measure spectra of diffuse reflectance, phosphorescence or fluorescence (Shopov, Georgiev, 1987, 1989) of the speleothem. It is designed for research of wideline spectra, such as luminescence of most speleothems formed at normal cave conditions (at temperature below 40° C), Shopov et al. (1989a). It allows easy non-destructive determination of objective information for mineral composition and speleothem luminescence, easy collection of information for cave minerals and conditions of their formation in caves. Conventional luminescent research methods have number of disadvantages, so several special speleothem research methods have been developed recently (table 2). They allow considerable enlargement of kinds and quality of the obtainable information.

Table 2. Special Speleothem Luminescence Research Methods

| Method | Authors | Obtainable information |
|---|---|---|
| I. Impulse Photography of Luminescence (IPL): 1. Photography of phosphorescence (IPP) 2. Photography of fluorescence & phosphorescence (IPFP). | Shopov & Tsankov (1986) Shopov & Grynberg (1985) | Diagnostics of minerals; registration of colour & zonality of fluorescence, phosphorescence and its spectra; UV photography; extraction of single mineral samples; chemical changes of the mineral-forming solution; Climate and Solar Activity variations during the Quaternary. |
| II. Laser Luminescent Microzonal Analysis (LLMZA) | Shopov (1987) | Microzonality of luminescence; changes of the mineral-forming conditions; High Resolution Records of Climate & Solar Activity variations (with resolution up to 0.4 days). Reconstruction of annual rainfall and annual temperature in the past. Estimation of past Cosmic Rays (CR) and Galactic CR. Speleothem growth hiatuses. |
| III. Color Slide Spectrophotometry (CSS) | Shopov & Georgiev (1989) | Wide-band spectra of phosphorescence, fluorescence and diffuse reflectance of minerals; spectra of quick processes. |
| IV. Autocalibration Dating (ACD) | Shopov, Dermendjiev & Buyukliev (1991) | High Precision Speleothem Dating of Speleothems of any age, Climatic and Solar Activity cycles, variations of the Speleothem Growth Rate. |
| V. Time Resolved Photography of Phosphorescence (TRPP) | Shopov et al.(1996-d) | Determination of the lifetime of the luminescent center. Uplift of the region. Past mixing of surface and epithermal or hydrothermal waters during mineral growth. Estimation of the temperature of the deposition, plus all information obtainable by IPP |

**Paleoenvironmental applications of Speleothem Luminescence**

Before using a speleothem for any luminescence paleoenvironmental records it is necessary to determine that all luminescence of the sample is due to organics. Otherwise a subsequent research may produce major confusions. To prove that all speleothem luminescence is due to organics is a very complicated task (Shopov, 1997).

**Paleoluminescence, Paleotemperature and Paleo-Solar Activity.**

Calcite speleothems frequently display luminescence, which is produced by calcium salts of humic, and fulvic acids derived from soils above the cave (Shopov, 1989a,b; White, Brennan, 1989). These acids are released (i) by the roots of living plants, and (ii) by the decomposition of dead matter. Root release is modulated by visible (650- 710 nm) solar insolation (SI) via photosynthesis, while rates of decomposition depend exponentially upon soil temperatures that are determined primarily by solar infrared radiation (Shopov et al., 1994) in case that the cave is covered only by grass or upon air temperatures in case that the cave is covered by forest or bush. In the first case, microzonality of luminescence of speleothems can be used as an indirect Solar Activity (SA) index (Shopov et al., 1990c), but in the second as a paleotemperature proxy. So in dependence on the cave site we may speak about "solar sensitive" or "temperature sensitive" luminescent speleothem records like in treering records, but in our case record may depend only on temperature either on solar irradiation:
- In the case of Cold Water cave, Iowa, US we obtained high correlation coefficient of 0.9 between the luminescence record and Solar Luminosity Sunspot index measured since 1700 AD (Shopov et al., 1996-a) and reconstructed sunspot numbers since 1000 AD with precision of 10 sunspot numbers (which is within the experimental error of their measurements);
- In the case of Rats Nest cave, Alberta, Canada we measured correlation of 0.67 between luminescence intensity and air temperatures record for the last 100 years and reconstructed annual air temperatures for last 1500 years at the cave site with precision of 0.35° C (Shopov et al., 1996-a). Intensity of luminescence was not dependent on actual precipitation and sunspot numbers (zero correlation).
Time series of a Solar Activity (SA) index "Microzonality of Luminescence of Speleothems" are obtained by Laser Luminescence Microzonal analysis (LLMZA) of cave flowstones described by Shopov (1987). This technique uses a relatively simple device, but the quality of results is as good as high is the experience of the researcher, because every sample requires a different approach. Many restrictions for samples for LLMZA apply (Shopov, 1987). LLMZA allow measurement of luminescence time series with duration of hundreds of thousands years, but time step for short time series can be as small as 6 hours (Shopov et al., 1994) allowing resolution of 3 days (Shopov, et al., 1988). IPP and LLMZA devices (Shopov & Grinberg, 1985, Shopov& Tsankov, 1986, Shopov, 1987) are the only ones allowing reliable measurements of the intensity of luminescence of speleothems. The wide range of devices used for measurement of speleothem annual growth by annual bands of luminescence do not produce reliable intensity of luminescence of speleothems, so can not be used for any other luminescent paleoenvironmental reconstructions.
Luminescence proxy records of the Solar Activity have been used for solving many astrophysical problems (Dermendjiev et al. 1989, 1990, 1992)

**Paleoluminescence Reconstructions of the Solar Insolation**
Basically all solar sensitive raw paleoluminescence records (if measured properly using IPP or LLMZA devices) are solar insolation records (Stoykova et al., 1998, Shopov, et al., 2000). Other proxies can be derived from such records using different types of digital analysis.

**Paleoluminescence Reconstructions of the Solar Luminosity**
NASA used a record of luminescence of a flowstone from Duhlata cave, Bulgaria to obtain a standard record of variations of the Solar Irradiance ("Solar constant") in [W/m$^2$] for the last 10000 years (D. Hoyt, personal communication) by calibration of the luminescence record of (Shopov et al., 1990b) with satellite measurements.
Paleoluminescence solar insolation proxy records contain not only orbital variations, but also solar luminosity self variations, producing many cycles with duration from several centuries to 11500 years with amplitude ranging respectively from 0.7 to 7 % of the Solar Constant (Shopov et.al 2004, same volume). Solar luminosity variations can be obtained from paleoluminescent records by extracting of the orbital variation from them using band- pass filtration with frequencies of the orbital variations.

These millennial solar luminosity cycles can produce climatic variations with intensity comparable to that of the orbital variations. Known decadal and even century solar cycles have negligible intensity (100 times less intensive) relatively to these cycles. Solar luminosity (SL) and orbital variations both cause variations of solar insolation affecting the climate by the same mechanism.

Luminescence time series have been used to solve a number of problems of solar physics (Dermendjiev et al., 1989, 1990, 1992).

**Luminescence and Cosmic Rays Flux (CRF)**

Cosmic rays produce cosmogenic isotopes ($^{14}$C, $^{10}$Be, etc.) in the upper atmosphere by nuclear reactions. As it is known, the $^{14}$C variations record represents the Cosmic Ray Flux (CRF) and modulation of the CRF by the solar wind (representing solar activity). We have obtained a striking correlation (with a correlation coefficient of 0.8) between the calibration residue delta $^{14}$C record and a luminescent speleothem record (Shopov et al., 1994). It is as high as the best correlation ever obtained between a direct Solar index (inverted annual Wolf number) and the CRF (Beer, 1991, r= 0.8). Obviously luminescence records can be used as a CRF proxy. To reconstruct the past CRF the luminescent record should be inverted. In this way a reconstruction of the solar modulation of the CRF during the last 50000 years with resolution of 28 yrs was obtained.

**Luminescence and supernova explosions**

Galactic CRF have some short- term variations due to supernova explosions. These variations of the GCRF can be determined only by comparison of a record of production of cosmogenic isotopes (CI) with an independent on CRF solar activity record. The luminescence microzonality is the only independent SA index with such length of record. It was used to reconstruct GCRF variations for the last 6500 years with 20-yr. resolution by subtracting of an inverted luminescent SA record from the residual 14C record (Shopov et al., 1996-b). Obtained record represents self-variations of the GCRF (due to supernova eruptions) beyond Solar system (where solar modulation does not exist).

**Luminescence Reconstructions of the Variations of Geomagnetic Field**

Variations of intensity of the Geomagnetic field dipole also correlates with speleothem luminescence (Shopov et al. 1996-e), because the geomagnetic field is modulated by the magnetic field of the solar wind (which is one appearance of the solar activity). This is due to formation of induced electromagnetic field in the Earth's magnetosphere in result of rotation of the Earth (which has own magnetic field) in the variable magnetic field of the Solar wind. This process is similar to rotation of a dynamo machine (electric generator).

**Luminescence and Paleosoils**

Luminescence organics first detected in speleothems by Gilson and Macarthney (1954) are humic and fulvic acids accordingly White, Brennan, 1989, but more precisely they can be divided to 4 types according to Shopov (1997):

1. Calcium salts of fulvic acids, soluble in water;
2. Calcium salts of humic acids not soluble in water and acids, but mobile in karst in form of colloid solutions;
3. Calcium salts of huminomelanic acids not soluble in waters and acids, but soluble in alcohols. They are mobile in karst in form of colloid solutions.
4. Organic esters not soluble in water but soluble in ether.

Each of these classes is usually presented in a single speleothem with hundreds of chemical compounds with similar chemical behavior, but different molecular weights. Superposition of luminescence bands of all these compounds gives the broadline spectra of luminescence of organics in a speleothem. Distribution of concentration of these compounds (and their luminescence spectra) depends on type of soils and plants society over the cave. So study of luminescence spectra of these organics can give information about paleosoils and plants in the past (White, Brennan, 1989). Changes in the visible colour of luminescence of speleothems suggest major changes of plants society and are observed very rarely only in speleothems growing hundreds of thousands years through glacial and interglacial periods (Shopov, 1997).

**Luminescence and Annual Growth Rates of Speleothems**

Speleothem growth rate may vary up to 14 times within a single sample, resulting in non-linear time scale of the records (Shopov, et al.1992, 1994). These variations represent rainfall variations if there are no growth interruptions (hiatuses) between the dating points in the speleothem. Speleothem luminescence visualizes annual microbanding, not visible in normal light (Shopov, 1987, Shopov et al., 1988a). This peculiar banding have been called "Shopov bands" by S.E. Lauritzen et al. (1996). If this banding is visible in normal light or the luminescent curve has sharp profiles or jumps like in Baker et al. (1993), it suggests that speleothem growth stopped for a certain period during the year and such time series can not be used for obtaining rainfall proxy records. Maxima of intensity of luminescence reflect air temperature in August, but minima in February in a speleothem from Rats Nest cave, Alberta, Canada (Shopov et al., 1996-a, c). By measuring the distance between subsequent maxima we may derive a proxy record of annual precipitation for the cave site. Shopov et al. (1996-a, c) measured correlation of 0.57 between speleothem annual growth rates (determined by measuring of the distance between maxima of the annual structure of luminescence of a speleothem from Rats Nest cave, Alberta) and the historic record of measured annual precipitation in Banff, Alberta (from August to August) for the last 100 years. This way was reconstructed annual precipitation for last 280 years at the cave site with precision of 80 mm/year (Shopov et al., 1996-a, c). Now studies of 15 labs over the world are concentrated only on this specific topic of paleoenvironmental applications of speleothem luminescence (Baker et al., 1993, Lauritzen et al., 1996, etc.)

We obtained a reconstruction of growth rates and precipitation (for the last 6400 years with averaged time step of 41 years) for Iowa (near Cold Water cave), US (Shopov, et al., 1996-c) by "tuning" of the time scale of a luminescent record to the calibration 14C record (Stuiver and Kra, 1986). It suggests higher speleothem growth rate and higher precipitation between 6400- 2500 years B.P. at the cave site. This speleothem is dated with 7 TIMS U/Th dates.

**Using of Paleoluminescence for Determination of the Origin of Glacial Periods and Improvement of their Dating**

We measured a luminescent solar insolation proxy record in a speleothem (JC11) from Jewel Cave, South Dakota. This record exhibits a very rapid increasing in solar insolation at 139 kyrs +/- 5.5 kyrs (2 sigma error) responsible for the termination II. This increasing is preceding the one suggested by the Orbital theory with about 10 kyrs and is due to the most powerful cycle of the solar luminosity with duration of 11.5 kyrs superposed on the orbital variations curve. Solar luminosity variations appear to be as powerful as orbital variations of the solar insolation and can produce climatic variations with intensity comparable to that of the orbital variations (Shopov et al. 2004). So paleoluminescence speleothem records may serve as a reliable tool for studying the mechanisms of formation and precise timing of glaciations.

**Paleoluminescence and Sea Level Variations**

Using speleothem luminescence solar insolation proxy records it has been demonstrated, that solar luminosity variations are responsible for almost 1/2 of the variations in high-resolution solar insolation experimental records. Solar luminosity variations are responsible for the short time variations of the sea level (Shopov et al. 2000).

**Luminescence Reconstructions of Past Karst Denudation**

Reconstructions of past carbonate denudation rates have been made using the quantitative theory (Shopov et. al, 1991) of solubility of karst rocks (in dependence of the temperature and other thermodynamic parameters) and quantitative paleoluminescence reconstructions of the annual precipitation rates (for the last 280 years) and of the annual temperature for the last 1200 years (Shopov et al. 2001).

**Pollution and migration of toxic compounds indicated by speleothem luminescence**

In many samples all or a significant part of the luminescence is produced by ions of uranium and Pb. Sometimes they even have annual banding due to variations of acidity of the karst waters (Shopov, 2004), causing variations of the solubility of some pollutants or toxic compounds (Shopov, 1997). Uranium compounds have such migration behavior.

**Luminescence of Hydrothermal Minerals**

Luminescence of the high- temperature hydrothermal minerals is due mainly to cations because molecular ions and molecules destruct at high temperatures. Luminescence of some cations can be used as an indicator of hydrothermal origin of the cave mineral (Shopov, 1989 a, b). Calcites formed by low-temperature hydrothermal solutions have short-life fluorescence due to cations and long phosphorescence of molecular ions (Gorobetz, 1981). Minimal temperature of appearance of this orange-red luminescence of calcite was estimated by Dublyansky (in press) to be about 40 °C by fluid inclusion analysis in hydrothermal cave calcites, but our direct measurements of luminescence of calcites in hot springs show that even at 46 °C such luminescence did not appear. It probably appears at over 60 °C. Such luminescence data are comparable with the stable isotope data used conventionally for this purpose (Bakalowicz et. al., 1987, Ford et al., 1993).

**Luminescence and Tectonics**
The tectonic uplift of an area (i.e., uplift of bedrock) can be deduced by luminescence in combination with absolute dating methods. For example some speleothems from Carlsbad Cavern, New Mexico exhibit luminescence originated by epithermal mineralizing solutions in the older part of the speleothem. Mixing of these waters with surface waters containing organics appear in younger parts of the speleothem, thus indicating the uplift during the duration of speleothem deposition (Shopov et al., 1996-d). This allows to date the uplift of the Gaudalupe Mountains. The boundary layer (so the uplift) can be dated by U/Pb dating methods (Ford, 2002).

**Luminescence and Dating of Speleothems**
Finally, speleothem's luminescence may be used to determine the age of the speleothem itself. Shopov et al. (1991a) and Dermendjiev et al. (1996) developed the new Autocalibration dating technique, which is shown to be the most precise speleothem dating method for samples younger than 2000 years (Shopov et.al, 1994).
Although Ugumori and Ikeya (1980) first suggested Optically Stimulated Luminescence (OSL) dating technique on speleothem calcite further attempts were not successful due to interference of luminescence of organics. Therefore OSL- dating cannot be used for speleothems.

**Conclusions**
In conclusion speleothem luminescence of organics can be used for obtaining of broad range of paleoenvironmental information. Most of the progress in this field was made in result of the operation of the International Programs "Luminescence of Cave Minerals" and "Speleothem Records of Environmental Changes in the Solar System" of the commission on Physical Chemistry and Hydrogeology of Karst of UIS of UNESCO.
Some speleothems can be used as natural climatic stations, for obtaining of proxy records of Quaternary climate with annual resolution.

**Acknowledgements**
This research was funded by Bulgarian Science Foundation by research grant 811/98 to Y. Shopov